# A new porous metallic silicon dicarbide for highly efficient Li-ion battery anode identified by targeted structure search

Junyi Liu,[a] Shuo Wang,[a] Yu Qie,[a] Jiabing, Yu,[a] Qiang Sun,[a,b, *]

[a] Department of Materials Science and Engineering, Peking University, Beijing 100871, China

[b] Center for Applied Physics and Technology, Peking University, Beijing 100871, China

**Abstract**

Extensive efforts have been devoted to C-Si compound materials for improving the limited specific capacity of graphite anode and avoiding the huge volume change of Si anode in Li-ion battery, but not much progress has been made during the past decades. Here, for the first time we apply the *targeted structure search* by using Li in *desired quantity* as chemical template to regulate the bonding between C and Si, which makes searching more feasible for us to find a new stable phase of $C_2Si$ (labelled as T-$C_2Si$) that can better fit the XRD data of silicon dicarbide synthesized before. Different from the conventional semiconducting silicon carbides, T-$C_2Si$ is not only *metallic* with high intrinsic conductivity for electrons transport, but also *porous* with regularly distributed channels in *suitable size* for Li ions experiencing a low energy barrier. T-$C_2Si$ exhibits a high specific capacity of 515 mA/g, a high average open-circuit voltage of 1.14 eV, and a low volume change of 1.6%. These parameters meet the requirements of an ideal anode material with high performance for electric vehicles. Moreover, our *targeted search* strategy guarantees the resulting anode material with a desirable specific capacity and a small volume change during charging /discharging, and it can be used to find new geometric configurations for other materials.

**TOC**

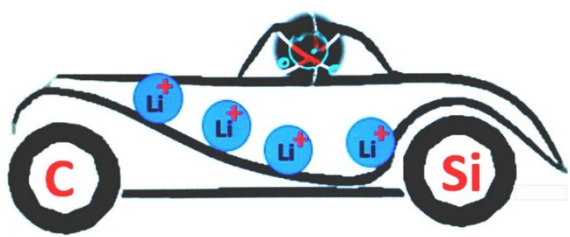

**Introduction**

Battery science and technology are of current interest. Conventional anode materials for Li-ion battery (LIB) are primarily based on graphitic carbon due to the merits of good stability, high electrical conductivity and low cost. However, the theoretical capacity is only 372 mAh/g that can't fulfil the requirements of electric vehicles pursued currently[1]. In contrast to graphitic carbon, Si-based materials have much higher specific capacities of 4,200 mAh/g[2], but they suffer from their own problems: (i) the poor reversibility and stability because of the huge volume change during intercalation and deintercalation of Li ions (>300%); (ii) poor rate capability because of the low electronic conductivity[3]. To ameliorate the electrochemical performances of Si-based anode, a variety of methods, such as reducing the size of silicon[4, 5], porous structures[6], and metal doping[7, 8], have been explored. Among them, introducing carbon into the silicon is considered as the most attractive strategy that can combine the merits of the carbon and Si. For example, Shi *et al.*[9] suggested that silicene/graphene heterostructure can not only retain the high lithium capacity of silicene and good electronic conductivity of graphene but also give rise to a much higher lithium binding energy via synergistic effect. Experimentally, a lot of Si–C composites like yolk–shell structured Si–C nanocomposites, 3D graphene–silicon networks[10, 11] are synthesized for the same reason. While, for these Si-C heterostructures and composites, there are still some problems to be solved in the practical applications including (i) non-conductive polymers are commonly adopted to bind Si and carbon nanoparticles together, thus reducing the capacity and electronic conductivity[12]; (ii) Si far away from the carbon structures is not able to participate in the battery reaction, and it is difficult to distribute the carbon and Si precisely[13]; (iii) There is no enough space to accommodate the volume expansion during the lithiation process, resulting in the fracture of the electrode structure after a few cycles[14]. Although using the nanocomposites can moderate the volume expansion effect, the nanoparticles and monolayer materials tend to cluster and restack duo to the high surface energy[15], which would not exhibit their inherent superiority in the practical application. Then a question arises: Can we find a 3D C-Si compound with intrinsic high electronic conductivity as well as ordered channels to avoid the problems discussed above?

Here we show that it is possible to find such C-Si compounds by using global structure search method, which is widely adopted to find the stable or metastable materials with specific properties. Such examples include Si with direct band gap[16], cathode materials of new $Li_2MnO_3$ allotropes[17], and lithium superionic conductor of oxysulfide LiAlSO[18]. Instead of searching for porous C-Si compounds *directly*, we adopt Li in *desired quantity* as synthesizing template to regulate the geometric configuration during forming Li-containing 3D porous C-Si compounds, and then get 3D porous C-Si structure by removing the Li atoms. By this means, we can guarantee that the material has enough space to accommodate Li ions and avoid large volume changes during lithiation and delithiation, we call this method *targeted structure search*. The feasibility of this strategy can be clearly seen from the recent success in synthesizing porous 3D $Si_{24}$ structure via $Na_4Si_{24}$ precursor by removing the sodium through a thermal degassing process[19]. In this study, we choose the chemical component of $LiC_2Si$ for the structure-searching with the following reasons: (i) $C_2Si$ is a stable neutral molecule in gaseous phase which has been widely found in the circumstellar envelope of carbon rich stars[20]; (ii) Quasi-two-dimensional $C_2Si$ has been fabricated successfully in experiment[21]; (iii) Silicon dicarbide crystal was synthesized in mixtures of hardened novolac resins with silicon power by thermal treatment up to 1200 °C[22], but its geometry has not yet convincingly determined, the previous theoretical work just used the structure of 3D glitter C doped with Si[23, 24]. The rich features of Si-C compounds[25] further stimulate us to explore $C_2Si$ mediated by Li. Through intensive crystal structure search, we have identified a new phase of $C_2Si$ compound that better fits the experimental XRD data, being porous and metallic with ordered nano channels in suitable size for the storage and transport of Li ions with a high capacity and fast kinetics.

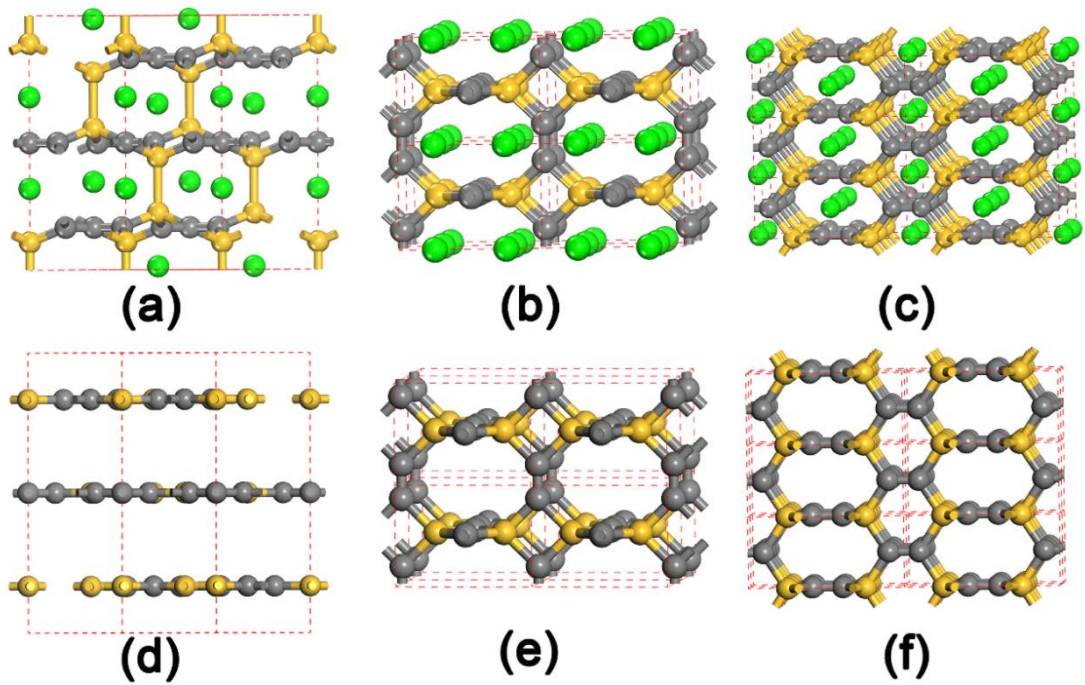

**Figure 1.** The structure of (a) L-LiC$_2$Si, (b) T-LiC$_2$Si, (c) G-LiC$_2$Si. The corresponding structure without Li are (d) L-C$_2$Si, (e) T-C$_2$Si and (f) G-C$_2$Si, respectively.

**Results and Discussion**

**Structure searching.** First, we perform the global structure search based on Particle Swarm Optimization (PSO) algorithm as implemented in the CALPSO code[26] to find all the possible 3D structures of LiC$_2$Si within a supercell containing 24 atoms (corresponding to six formulae) with lower energy and porous C$_2$Si frame, three of which are shown in the Fig. 1 (a-c). Fig. 1(a) is the ground state with the lowest energy, where C and Si atoms form a layered honeycomb structure and the layers are connected through Si-Si bonds, therefore, we label it as L-LiC$_2$Si. Fig. 1(b) is the second stable structure with 0.098 eV/f.u. higher in energy, where all the carbon dimers are bonded with the tetra-coordinated silicon and form into tetragons, and we label it as the T-LiC$_2$Si. Fig. 1(c) is a similar structure with Fig. 1(b) with 0.73 eV/f.u. higher than that of the ground state, differing from structure *b,* the carbon dimers and Si atoms form into hexagons not tetragons. In addition, the C$_2$Si frame of Fig. 1(c) is in the form of glitter structure which is reported in the previous study[24], therefore, we label the structure in Fig. 1(c) as G-LiC$_2$Si. Then we remove the Li atoms from the three structures to get the porous C$_2$Si structures. After fully relaxation, the resulting C$_2$Si structures are presented in the Fig. 1(d-f) and labeled as L-C$_2$Si, T-C$_2$Si and G-C$_2$Si, respectively. The Lattice parameters of these structures are presented in the Table 1. For the L-LiC$_2$Si, the frame of C$_2$Si changes to layered structure with broken Si-Si bonds, suggesting the structural phase transition during lithiation and delithiation. For the G-LiC$_2$Si, the lattice show a 10% contraction along c direction and 4.3% expansion along a/b direction. While for T-LiC$_2$Si, the frame of C$_2$Si shows slight changes after the Li atoms being removed, the total volume change is only 1.6%, which is comparable to that of the "zero-strain material" Li$_4$Ti$_5$O$_{12}$[27] and can be expected have an excellent cycle life. The big difference in volume changes between G-LiC$_2$Si, and T-LiC$_2$Si is due to their different geometries. When comparing the size of large channels, the shortest distance between C$_2$ –C$_2$ in G-C$_2$Si is only 3.18 Å that is less than the value of graphite (3.35 Å), while the corresponding value for T-C$_2$Si is 4.02 Å. This clearly indicates that G-C$_2$Si is not suitable for lithium battery anode.

To determine the geometric configuration of experimentally synthesized C$_2$Si sample[22], we have calculated XRD and compared with experimental data as shown in Fig. 2, which clearly indicates that T-C$_2$Si phase better fits the data, while in previous studies only G-LiC$_2$Si phase was considered[22] by simply replacing C with Si on sp$^3$ sites in glitter C structure. Based on above discussions, we focus our study on T-C$_2$Si phase to explore its possibility as high-performance anode.

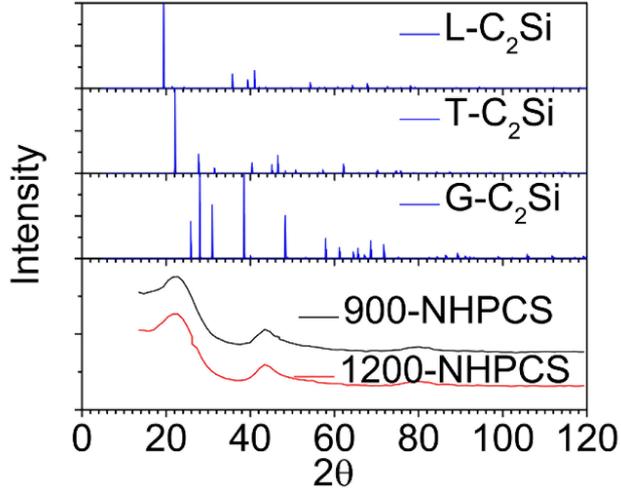

**Figure 2.** The comparison of simulated XRD with experimental data (the bottom panel).

**Table 1.** Lattice parameters and atomic positions of L-LiC$_2$Si, T-LiC$_2$Si, G-LiC$_2$Si and the corresponding L-C$_2$Si, T-C$_2$Si, G-C$_2$Si, respectively.

| structures | Space group | Lattice parameters (Å) | Atomic positions(fractional) |
|---|---|---|---|
| L-LiC$_2$Si | *R32*(155) | a=b=4.976, c=11.482 | Li 6c (0.000 0.000 0.820)<br>Si 6c (0.000 0.000 0.391)<br>C 3a (0.000 0.000 0.000)<br>C 9d (0.707 0.000 0.000) |
| T-LiC$_2$Si | *P4$_2$/mmc*(131) | a=b=4.014, c=5.466 | Li 2e (0.000 0.000 0.250)<br>C 4m (0.176 0.500 0.000)<br>Si 2f (0.500 0.500 0.250) |
| G-LiC$_2$Si | *P4$_2$/mmc*(131) | a=b=3.051, c=7.635 | Li 2c (0.000 0.500 0.000)<br>C 4i (0.000 0.500 0.404)<br>Si 2e (0.000 0.000 0.250) |
| L-C$_2$Si | *R32*(155) | a=b=5.018, c=13.707 | Si 6c (0.000 0.00 0.000)<br>C 3a (0.000 0.000 0.000)<br>C 9d (0.712 0.000 1.000) |
| T-C$_2$Si | *P4$_2$/mmc*(131) | a=b=4.017, c=5.368 | C 4m (0.167 0.500 0.000)<br>Si 2f (0.500 0.500 0.250) |
| G-C$_2$Si | *P4$_2$/mmc*(131) | a=b=3.182, c=6.884 | C 4i (0.000 0.500 0.4007)<br>Si 2e (0.000 0.000 0.250) |

**Stability.** Before examining the performance of T-C$_2$Si as an anode, we need to confirm the stability of T-C$_2$Si and its fully Li-inserted states T-LiC$_2$Si. To this end, we perform the phonon spectrums calculations and ab initio molecular dynamics (AIMD) simulations. As shown in the Fig. 3 (a-b), the phonon spectrums suggest that both T-C$_2$Si and T-LiC$_2$Si are dynamical stable as all of their vibration modes are real in their entire Brillouin zones. The AIMD simulations at the 300 K are performed within 4 × 4× 3 supercell for T-C$_2$Si and T-LiC$_2$Si. The both structures show no irreversible change during the AIMD simulations and the corresponding potential energies, presented in Figure 3 (c-d), only fluctuate around certain constant magnitude. Thus, we can conclude that T-C$_2$Si and T-LiC$_2$Si are metastable and the energy barrier of phase change are large enough against the thermal fluctuations at room temperature. In addition, we also examine the mechanical stability of T-LiC$_2$Si and T-C$_2$Si by calculating their linear elastic constants. Crystals of the tetragonal class have six independent elastic constants, and the corresponding stability criteria are as follow[28]:

$$C_{11} > |C_{12}|, C_{33}(C_{11} + C_{12}) > 2C_{13}^2 \qquad (1)$$

$$C_{44} > 0, C_{66} > 0 \qquad (2)$$

It is obvious that elastic constants of T-C$_2$Si and T-LiC$_2$Si, presented in the Table 2, satisfy those conditions, indicating that T-C$_2$Si and its fully Li-intercalated states T-LiC$_2$Si are mechanically stable and are able to stand against certain mechanical shocks during the battery cycles.

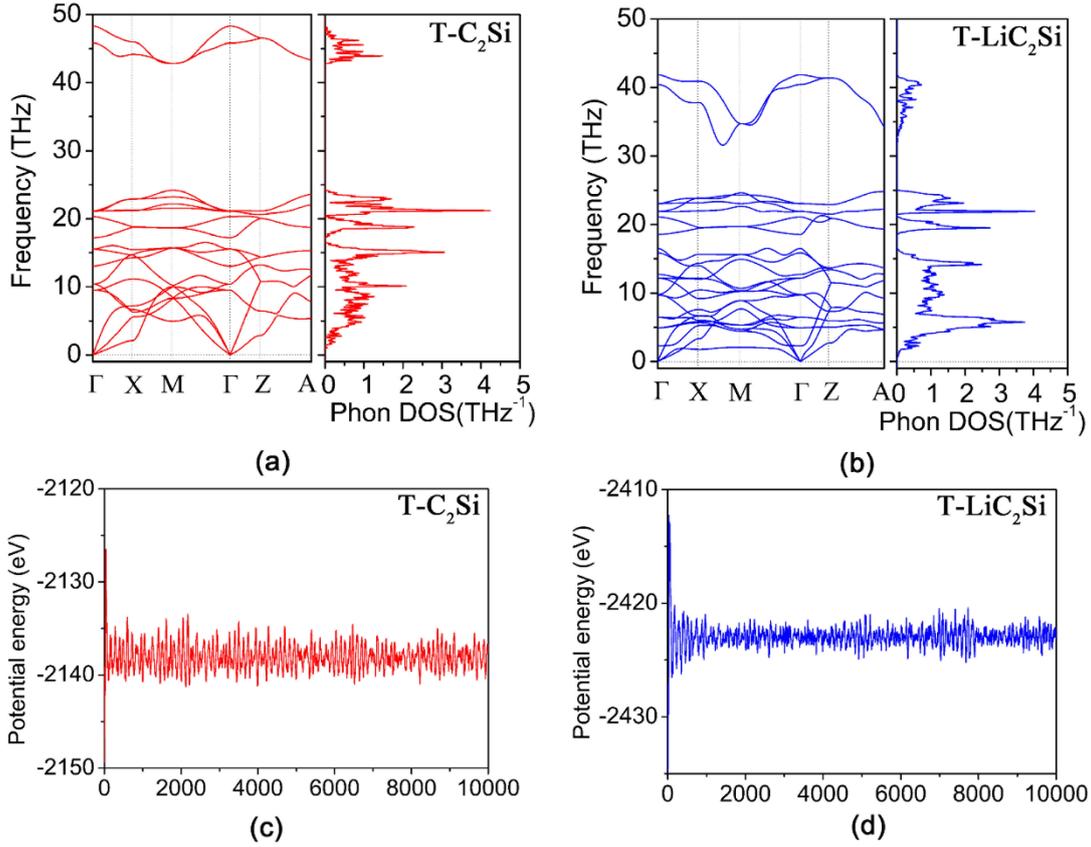

**Figure 3.** Stability of T-C$_2$Si and T-LiC$_2$Si. Phonon spectra and the frequency density of states of (a) T-C$_2$Si and T-LiC$_2$Si. The fluctuation of potential energy with respect to AIMD simulation with a time step of 1 fs for (c) T-C$_2$Si and T-LiC$_2$Si (d) at 300 K.

**Table 2.** Elastic constants(C) of T-C$_2$Si and T-LiC$_2$Si.

|  | $C_{11}$(GPa) | $C_{12}$ | $C_{13}$ | $C_{33}$ | $C_{44}$ | $C_{66}$ |
|---|---|---|---|---|---|---|
| T-C$_2$Si | 336.6 | 10.8 | 69.9 | 284.3 | 19.7 | 4.5 |
| T-LiC$_2$Si | 335.3 | 49.8 | 63.1 | 368.4 | 23.6 | 21.2 |

**Electronic properties.** Electronic structure is a crucial property for anodes, which determines the conductive performance and reaction rate. In order to understand the electronic properties of the system, we plot the band structure of T-C$_2$Si and T-LiC$_2$Si in Fig. 4. Clearly, the T-C$_2$Si and T-LiC$_2$Si are both metallic. Thus, the new predicted T-C$_2$Si would pose a high electronic conductivity and show a fast charging/discharging response as anode. This intrinsic electronic characteristic would reduce the usage of conductive additive, which can avoid the extra loss of the anode capacity and lower the cost.

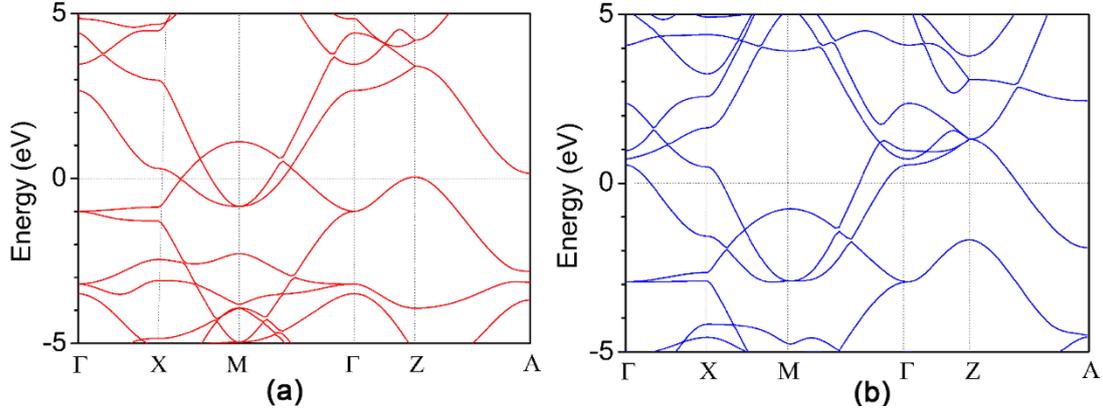

**Figure 4.** The band structures of (a) T-$C_2Si$ and (b) T-$LiC_2Si$

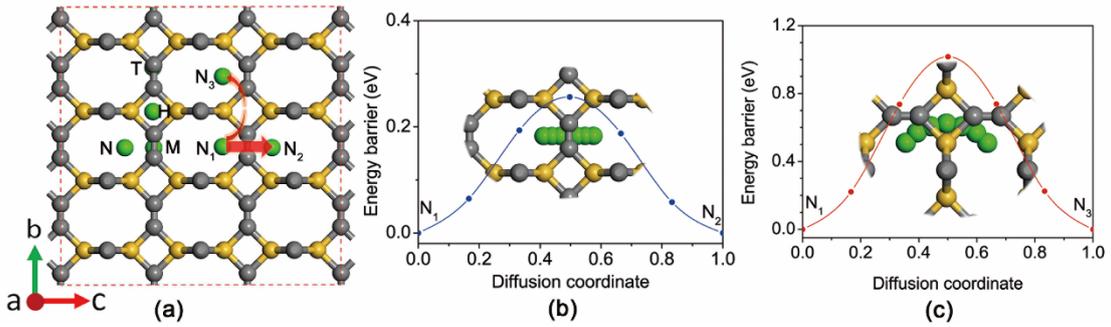

**Figure 5.** (a) The schematics of the C path ($N_1$–$N_2$) and AB path ($N_1$–$N_3$) in a side view, where $N_1$ is the initial point, and $N_2$, $N_3$ are the end points. (b) and (c) are the corresponding energy profiles.

**Single ion adsorption and diffusion in T-$C_2Si$.** Then we explore the adsorption and diffusion of single Li ion in T-$C_2Si$ to evaluate the performance of this material as an anode at low Li concentration. As seen from the Fig. 5, a 4×4×3 supercell of T-$C_2Si$ is adopted to avoid the interaction between the Li atoms, and four possible initial adsorption sites are considered: the neighboring site N (at the middle of two Si atoms and next to the C–C dimer), the bridge site M (above the midpoint of C-C bond), the hollow site H (above the center of the C-Si quadrangle), and the top site T ( above the top of the C atom). After relax the structure to the fullest, Li atoms on the different initial adsorption sites all converge to the N site, suggesting that N site is the most favorable adsorption site. For the adsorbed Li atoms, there is about 0.87 |e| charge transfer from Li to T-$C_2$Si based on the Bader charge population analysis, indicating that Li atoms donate almost all their valence electrons and turn into Li ions. Then we calculate the Li binding energy ($E_b$) to estimate the ionic binding with the matrix according to the following equation:

$$E_b = \frac{E_{T-Li_xC_2Si} - E_{T-C_2Si} - x\mu_{Li}}{x} \quad (3)$$

Where $E_{T-Li_xC_2Si}$ and $E_{T-C_2Si}$ are the total energies of Li-inserted and pristine T-$C_2$Si crystal structure, respectively. $\mu_{Li}$ is the chemical potential of Li which is taken as the cohesive energy per atom of metallic Li. The calculated $E_b$ of Li adsorption on N site of T-$C_2$Si is -1.68 eV, the absolute value of which is much larger than that of graphite (-0.78 eV in our computations with the same method), indicating a strong ionic binding between Li and the T-$C_2$Si.

Next, we investigate the Li-ion diffusion in T-$C_2$Si to further evaluate the rate performance. As the diffusion paths along the *a* and *b* directions are the same in consideration of the crystal symmetry, we only need to explore the diffusion barriers of two different migration paths: one is path $N_1$–$N_2$, which goes across the C-C bond along the c direction (C path); the other is path $N_1$–$N_3$, which goes over the tetragon along the a/b direction (AB path), as shown in Fig. 5(a). The calculated energy profiles along the C path and AB path are shown in Fig. 5 (b) and (c), respectively. For the C path, the energy barrier is 0.26 eV which is comparable to that of graphite (0.218–0.4 eV)[29-31] and smaller than that of commercially used anode materials based on $TiO_2$ (0.35–0.65 eV)[32-34], indicating that Li-ions diffusion along the C path is easy. While for the case of AB path, the energy barrier is rather large with a value of 1.02 eV. According to the transition state theory, the temperature-dependent diffusion constant (D) can be estimated by the Arrhenius equation[35]:

$$D = A\exp(\frac{-E_a}{K_BT}) \quad (4)$$

where $E_a$, $K_B$ and T are the energy barrier, Boltzmann's constant and the environmental temperature, and A is the pre-exponential factor. According to Eq. 4, the diffusion constant at 300 K along the C path is about $6.21 \times 10^{12}$ times larger than that of AB path, suggesting that Li-ion diffusion in the T-$C_2$Si exhibits a significant one-dimensional feature due to the much larger energy barrier along a and b direction; this is similar to that of 2D black phosphorus[36] and 3D titanium niobate[37] and Bco-$C_{16}$[38]. On the basis of the basic theory of diffusion, diffusion constant is targetedly proportional to the spatial dimension[39], i.e. $D \propto \frac{1}{2n}$, where n is the dimensionality of diffusion space. When the material has a consistent orientation, diffusion in 1D will give a better rate performance.

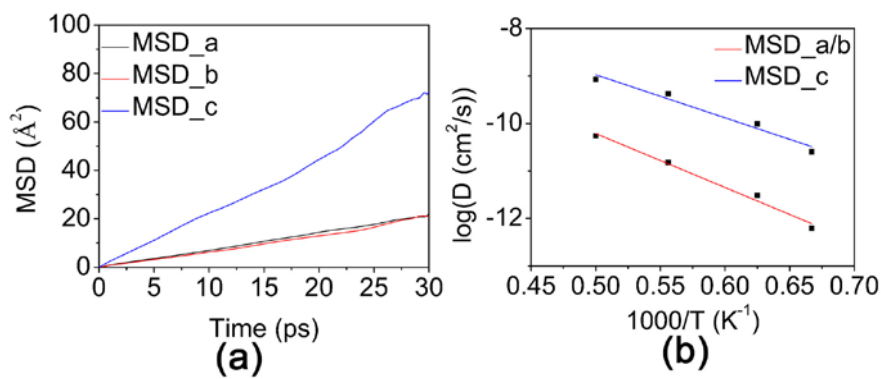

**Figure 6.** (a) The MSD of Li along different directions as a function of time at 2000 K with one Li vacancy in a 4x4x3 supercell. (b) Li ion diffusion constant along a/b direction (MSD_a/b) and c direction (MSD_c) as derived from mean square displacements of AIMD simulations at different temperature. The line is the corrsponding linear regression Arrhenius fit of the data.

**Theoretical Capacity and Vacancy Diffusion** Next, we investigate the fully Li-intercalated T-$C_2$Si which directly determines the theoretical Li capacity. The binding energy $E_b$ is calculated as −1.14 eV; the absolute value is larger than that of graphite (LiC$_6$: −0.11 eV) and close to that of VS$_2$ monolayer (Li$_2$VS$_2$: −0.93 eV)[40], suggesting that Li ions can be adsorbed safely in T-LiC$_2$Si and the clustering of Li ions would not happen at such a high Li concentration. The theoretical specific capacity of T-$C_2$Si is 515 mAh/g, which is about 1.4 times larger than that of graphite.

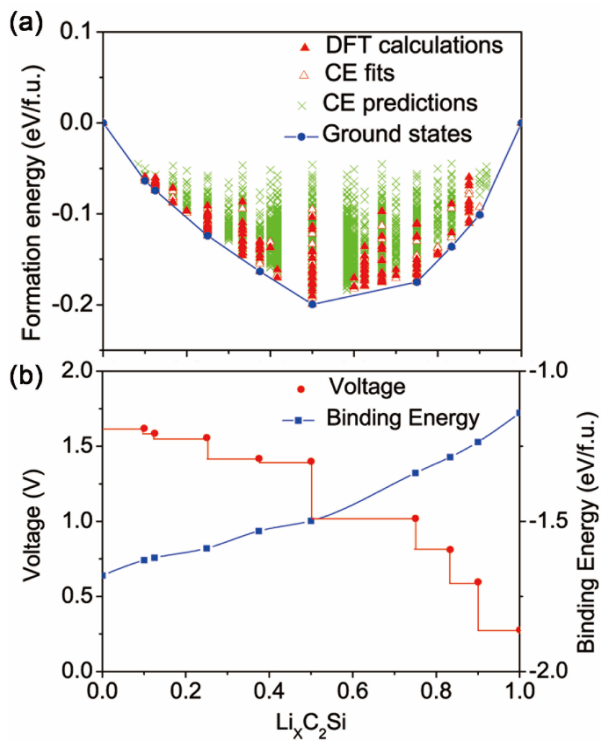

**Figure 7.** (a) Formation energies predicted by CE method based on the 141 different Li configurations with eight stable intermediate phases. (b) Corresponding voltage profile (marked in red) and binding energy profile (marked in blue) calculated by the eight stable intermediate phases.

To get a more complete picture of the Li ions diffusion in the T-LiC$_2$Si, we evaluate the diffusion at high Li concentration based on the AIMD simulation in a 4x4x3 supercell of T-LiC$_2$Si with one Li vacancy. The elevated temperature is adopted during the simulation to accelerate the migration process. The calculated mean square displacement (MSD) of Li ions at 2000 K is shown in Fig. 6, and the other MSD results at 1500, 1600, 1800 K are presented in the SI. Obviously, the MSD results are the same along $a$ and $b$ direction and the values are about 1/3 of that along c direction, suggesting that Li-ion diffusion still prefers the $c$ direction at high Li concentration. The Li-ion diffusion constants along different directions at different temperature are calculated, and the detailed results are given in the supporting information. Based on the diffusion constants, the diffusion energy barrier along $a/b$ and $c$ directions are estimated via linear regression Arrhenius fitting, as shown in the Fig. 6(b). The corresponding energy barrier along $a/b$ direction is 0.98 eV and 0.78 eV along $c$ direction, indicating that the Li concentration mainly affects the Li-ion diffusion along *c* direction because the increased repulsion and hindrance between Li ions while it has little influence on *a/b* direction. The theoretical extrapolated diffusion constants at room temperature are computed to be $5.58 \times 10^{-19}$ cm$^2$/s along a/b direction and $1.186 \times 10^{-15}$ cm$^2$/s along the c direction.

**Table 3.** Comparison of specific capacity, diffusion barrier, average open-circuit voltage (OCV) and volume change of candidate anode materials for Li Ion Battery. "-" means data unavailable.

| Materials | Specific capacity (mAh/g) | Diffusion barrier (eV) | OCV (eV) | Volume change |
|---|---|---|---|---|
| T-C$_2$Si | 515 | 0.26-0.78 | 1.14 | 1.6% |
| graphite | 372 | 0.218-0.4[29-31] | 0.11[40] | 10% |
| Li$_4$Ti$_5$O$_{12}$ | 175[27] | 0.13-0.7[41] | 1.55[27] | <0.2%[27] |
| TiO$_2$ | 200[42] | 0.3[43] | 1.8[42] | - |
| Bco-C$_{16}$ | 558[38] | 0.019-0.53[38] | 0.23[38] | 13.4%[38] |
| 2D black P | 432[44] | 0.08[36] | 2.9[36] | - |
| 2D MoS$_2$ | 335[40] | 0.25[40] | 0.26[40] | - |
| 2D VS$_2$ | 466[40] | 0.22[40] | 0.93[40] | - |
| 2D Ti$_3$C$_2$ | 447.8[45] | 0.068[45] | 0.43[45] | - |
| 2D Mo$_2$C | 526[46] | 0.043[46] | 0.68[46] | - |

**Theoretical Voltage Profile.** Open-circuit voltage is another important indicator that are widely used for charactering the performance of the Li battery, which is often obtained by calculating the energy difference over parts of the Li composition domain. Based on the general half-cell reaction vs. Li/Li$^+$, and in no consideration of the volume, pressure, and entropy effects, the average voltage of T-Li$_x$C$_2$Si in the concentration range of $x_1 < x < x_2$ can be calculated as follow[47]:

$$V \approx \frac{E_{T-Li_{x_1}C_2Si} - E_{T-Li_{x_2}C_2Si} + (x_2 - x_1)E_{Li}}{(x_2 - x_1)} \quad (5)$$

where $E_{T-Li_{x_1}C_2Si}$, $E_{T-Li_{x_2}C_2Si}$ and $E_{Li}$ are the total energy of $T-Li_{x_1}C_2Si$, $T-Li_{x_1}C_2Si$,

and metallic Li, respectively.

Based on the above formula, we need to search for the Li occupying configurations with lowest energy at different Li concentrations before calculating V. To this end, we construct a cluster-expansion (CE) Hamiltonian of T-Li$_x$C$_2$Si by fitting to the accurate first principles energies of 141 configurations with different Li concentrations. The fitting is carried out via Alloy Theoretic Automated Toolkit (ATAT) code[48] with a cross-validation (CV) score of 13 meV/f.u., which can ensure that the obtained CE Hamiltonian is accurate enough to predict the energies of any Li occupying configurations of T-Li$_x$C$_2$Si. The formation energy for a given Li-vacancy arrangement with composition x in T-Li$_x$C$_2$Si is defined as

$$E_f(\sigma) = E_{T-Li_xC_2Si} - (1-x)E_{T-C_2Si} - xE_{T-LiC_2Si} \quad (6)$$

which reflects the relative stability of the specific T-Li$_x$C$_2$Si structure with respect to phase separation into a fraction x of T-LiC$_2$Si and a fraction (1-x) of T-C$_2$Si. Based on the CE Hamiltonian, we calculate the formation energies of all the symmetry-inequivalent T-Li$_x$C$_2$Si structures within 60 atoms per cell and find eight stable configurations at intermediate concentrations, as shown in Fig. 7 (a). The corresponding concentration x (T-Li$_x$C$_2$Si) are 0.1, 0.125, 0.25, 0.375, 0.5, 0.75, 0.83 and 0.9, respectively. The corresponding configurations are presented in Fig. SI in the supporting information. The binding energies of these stable configurations are calculated and all are negative, suggesting that the Li ions can be adsorbed stably instead of clustering as Li metal. On the other hand, the absolute value decreases gradually with the increasing Li concentration duo to the enhanced repulsive interaction between Li ions. Using these stable configurations, we calculate the average open-circuit voltage based on the Eq. 5. The open-circuit voltage curve, in the Fig. 7(b), exhibits two prominent voltage regions: 0 < x < 0.5, it is a plateau region which corresponds to a weak interaction between the Li ions duo to the low Li concentration; 0.5 < x < 0.1, there is a dramatic drop from 1.39 to 0.28 V, which reflects an increasing repulsive interaction between the Li ions; By numerically averaging the voltage profile, the average voltage is found to be 1.14 V, which is comparable to that of 2D $VS_2$ (0.93 V)[40], and smaller than that of $TiO_2$ (1.8 V)[42], $Li_4Ti_5O_{12}$ (1.5 V)[27], and 2D black P (1.8-2.9 V)[36, 44], but much larger than that of graphite (0.11 V)[40] and 2D $Mo_2C$ (0.68 V)[46]. The higher average voltage can avoid the security problems caused by lithium dendrites, which is very important for the practical applications in electric vehicles.

**Summary**

To provide reliable theoretical clues for experimentalists to synthesize new anode materials going beyond the currently used ones, we have carried out comprehensive simulations combining DFT with global structure search and cluster expansion method. Different from the conventional studies on this subject reported so far, we have adopted *targeted structure search*. Instead of searching for C-Si structures *directly*, we use Li *in desired quantity* as synthesizing template to regulate the geometric configuration of Si-C, which provides us a flexibility to find a new structure T-$C_2Si$ with XRD better fitting the experimental results, suggesting the presence of T-$C_2Si$ in the synthesized silicon dicarbide reported before; Moreover, T-$C_2Si$ is more suitable for storing and transporting Li ions than G-$C_2Si$ phase, the new T-$C_2Si$ phase not only exhibits metallic features in band structure with a high intrinsic electronic conductivity but also displays regularly distributed channels with suitable size for Li ions to transport fast with a low energy barrier, the specific capacity is 515 mA/g, and the average open-circuit voltage is 1.14 V, while the channel size in the previously reported G-$C_2Si$ phase is less than that of graphite, so it is not good for anode material application; Equally important, our structure search method also suggests a new strategy to synthesize silicon dicarbide by using Li as synthesizing template in *ambient condition*, which is different from the method of thermal treatment up to 1200 degrees reported before. Moreover, our *targeted search* strategy can guarantee that the resulting anode material not only has a desirable specific capacity and reversible capacity but also has small volume changes during charging /discharging. In short, our newly identified T-$C_2Si$ phase not only better fits the experimental XRD data but also exhibits desirable features as a new anode material that improves the limited specific capacity of graphite and avoids the huge volume change of pure Si. We hope that this theoretical study would stimulate more effort to use silicon carbide anode materials for battery technology.

**Methods**

All the First-principles calculations are performed in the framework of density functional theory (DFT) as implemented in the Vienna ab-initio simulation package (VASP)[49]. The generalized gradient approximation (GGA) within the Perdew–Burke–Ernzerhof (PBE)[50] formalism is adopted for the exchange–correlation functional. The projector augmented wave (PAW) method[51] with a kinetic energy cutoff of 550 eV is used to expand the valance electron wave functions. Based on the convergence test, Brillouin-zone integration is performed in a regular k-mesh generated by the Monkhorst-Pack scheme[52] with a grid density of $2\pi\times0.02$ Å$^{-1}$. The convergence criteria of total energy and force components for self-consistent calculations and structure optimization are $1\times10^{-6}$ eV and 0.001 eV/Å, respectively. The calculation of Phonon spectrums are on the basis of finite displacement method implemented in the Phonopy code[53]. The diffusion barriers for Li ions are calculated based on the climbing-image nudge elastic band (CI-NEB) method[54,55] implemented in the VASP transition state tools, and the convergence criterion of force is 0.01 eV/Å. The detailed description of the structure search, MSD calculations and cluster expansion method are provided in the supporting information.

**AUTHOR INFORMATION**

Corresponding Author: Qiang Sun, sunqiang@pku.edu.cn

**ORCID**

Qiang Sun: **0000-0003-3872-7267**

**Notes:** The authors declare no competing financial interest.

**ACKNOWLEDGMENTS:** This work is partially supported by grants from the National Key Research and Development Program of China (2016YFB0100200), and from the National Natural Science Foundation of China (21573008 and 21773003). The calculations are supported by High-performance Computing Platform of Peking University.